\documentclass[iop]{emulateapj}
\usepackage{xspace,graphics,graphicx}

\newcommand\teff{\ensuremath{T_{\rm eff}}\xspace}

\slugcomment{Accepted for publication in the Astrophysical Journal} 
\shorttitle{Variable C-atmosphere White Dwarfs}
\shortauthors{Williams et al.}

\begin{document}
\title{Variability in Hot Carbon-Dominated Atmosphere (hot DQ) White Dwarfs: Rapid Rotation?}

\author{Kurtis A.~Williams\altaffilmark{1}, M.~H.~Montgomery\altaffilmark{2}, D.~E.~Winget\altaffilmark{2}, Ross E.~Falcon\altaffilmark{2,3}, and Michael Bierwagen\altaffilmark{4}
}
\email{Kurtis.Williams@tamuc.edu}
\altaffiltext{1}{Department of Physics \& Astrophysics, Texas A\&M University-Commerce, P.O. Box 3011, Commerce, TX, 75429, USA}
\altaffiltext{2}{Department of Astronomy, University of Texas, 1 University Station C1400, Austin, TX, 78712, USA}
\altaffiltext{3}{Sandia National Laboratories, Albuquerque, NM 87185-1196, USA}
\altaffiltext{4}{REU participant, Department of Physics \& Astrophysics, Texas A\&M University-Commerce}

\begin{abstract}
Hot white dwarfs with carbon-dominated atmospheres (hot DQs) are a cryptic class of white dwarfs.  In addition to their deficiency of hydrogen and helium, most of these stars are highly magnetic, and a large fraction vary in luminosity.  This variability has been ascribed to nonradial pulsations, but increasing data call this explanation into question.  We present studies of short-term variability in seven hot DQ white dwarfs.  Three (SDSS J1426+5752, SDSS J2200$-$0741, and SDSS J2348$-$0942) were known to be variable.  Their photometric modulations are coherent over at least two years, and we find no evidence for variability at frequencies that are not harmonics.  We present the first time-series photometry for three additional hot DQs (SDSS J0236$-$0734, SDSS J1402+3818, and SDSS J1615+4543); none are observed to vary, but the signal-to-noise is low.  Finally, we present high speed photometry for SDSS J0005$-$1002, known to exhibit a 2.1 d photometric variation; we do not observe any short-term variability.  Monoperiodicity is rare among pulsating white dwarfs, so we contemplate whether the photometric variability is due to rotation rather than pulsations; similar hypotheses have been raised by other researchers.  If the variability is due to rotation, then hot DQ white dwarfs as a class contain many rapid rotators.  Given the lack of companions to these stars, the origin of any fast rotation is unclear -- both massive progenitor stars and double degenerate merger remnants are possibilities.  We end with suggestions on future work that would best clarify the nature of these rare, intriguing objects.
\end{abstract}

\section{Introduction}
The Sloan Digital Sky Survey (SDSS) uncovered a spectral class of white dwarf (WD) exhibiting atomic carbon and/or oxygen absorption lines \citep{2003AJ....126.2521L}.  Shortly thereafter, the subset of these WDs with \ion{C}{2} in their spectra were recognized to be a new class of WD harboring carbon-dominated atmospheres, along with the previously known classes of WDs with hydrogen- and helium-dominated atmospheres \citep{2007Natur.450..522D,2008ApJ...683..978D}.  The origin of the high carbon abundances in these atmospheres, now known as spectral type ``hot DQ" due to their relatively high effective temperatures, was unknown.  The greatly enhanced carbon abundance of hot DQs is similar to that observed in the massive PG1159 stars H1509+65 and \objectname[RX J0439.8-6809]{RX J0439.8$-$6809} \citep{1991A&A...251..147W,2015A&A...584A..19W} and the ``second sequence" of cool DQ WDs \citep{2005ApJ...627..404D,2006A&A...454..951K}.

Spurred by a desire to better understand these mysterious hot DQs and by a simplistic model suggesting that hot DQs may be unstable to nonradial pulsations, we undertook a search for photometric variability in hot DQs, culminating in the discovery of $\sim 1.5\%$ amplitude modulations in the hot DQ SDSS J1426+5752\footnote{Long format identifiers for observed WDs are given in Tab.~ \ref{tab.obslog}} \citep{2008IBVS.5900....8D,2008ApJ...678L..51M}.  Our immediate assumption was that these modulations were nonradial pulsations, though we noticed that the pulse shape was very peculiar compared to other pulsating WDs.  We therefore posited that the moduations could be due to an accreting system such as \object{AM CVn}, though the lack of any additional evidence for accretion from any other spectral or photometric evidence has since led us and others to reject this hypothesis \citep{2008ApJ...683L.167D,2009ApJ...702.1593G,2013ApJ...769..123W}.

Subsequent to our discovery, variability was uncovered in several other hot DQs \citep{2008ApJ...688L..95B,2010ApJ...720L.159D,2011ApJ...733L..19D}.  Fascinatingly, the majority of these objects exhibit only a single independent period of modulation, though often with a significant harmonic, while most pulsating WDs show multiple independent modes of oscillation.  However, claims of low-amplitude non-harmonically related modes of variability \citep{2009ApJ...703..240D,2011ApJ...733L..19D,2009ApJ...702.1593G}, if verified, would resolve that seeming inconsistency.  

Meanwhile, other mysteries began to surface surrounding hot DQs and the seemingly new class of variable stars we called the DQVs.  At least 70\% of hot DQs have strong magnetic fields \citep[$\sim 1$ MG or greater;][]{2011ApJ...733L..19D,2013ASPC..469..167D}, a magnitude greater than the incidence of magnetism in the field.  Prior to SDSS J1426+5752, no strongly magnetic WD had been observed to pulsate \citep{2008ApJ...683L.167D}, as these magnetic fields should be sufficient to inhibit the fluid motions necessary for pulsations \citep[e.g.,][]{2015ApJ...812...19T} or shift the pulsations to high $l$ modes where they are difficult to detect.

In \citet{2013ApJ...769..123W}, we announced the discovery of short-period variability in a warm DQ WD, SDSS J1036+6522\footnote{Long format name: \object{SDSS J103655.39+652252.2}}.  The \teff of this WD, $15,500$ K, is well outside the instability strips calculated for hot DQ stars \citep[e.g.,][]{2008ApJ...678L..51M,2008A&A...483L...1F,2009A&A...506..835C}. In a separate study, \citet{2013MNRAS.433.1599L} announced the discovery of variability in the hot DQ  SDSS J0005-1002, except that this variability has a period of 2.1 d, far too long for nonradial pulsations in WDs and much similar to known periods of rotating WDs.  Are there multiple mechanisms for photometric variability among the hot (and warm) DQs, or have we not yet properly identified the mechanism behind the variations?

In order to better understand the variability in hot DQs, we undertook a multi-pronged approach.  First, we began a long-term observational study of three known variable hot DQs.  Rotation and coherent pulsations should maintain nearly constant period and phases over time, with changes possible over several years due to magnetic braking or WD evolution.  Additionally, very low amplitude coherent modulations should become detectable as the signal-to-noise increases in frequency space.  Second, we observed multiple hot DQs without previous time-series observations to search for coherent luminosity variations, primarily to determine the ubiquitousness of variability among hot DQs.

\section{Observations and Reduction}

We obtained our data over several long observing runs between the spring of 2008 and the summer of 2011 with the Argos high speed photometer on the McDonald Observatory's 2.1 m Otto Struve Telescope.   Argos contained a $512\times512$ pixel back-illuminated frame transfer CCD with a field of view of 2\farcm8 on a side \citep{2004ApJ...605..846N}.  Our observations were obtained through a 1 mm Schott glass BG40 filter.  Exposure times varied from 10 s to 30 s, depending on the apparent magnitude of the target, the observing conditions, and the observer's preference.  Due to the frame-transfer time of only 310 $\mu$s,  there is no significant loss of time to readout.  Precise timing was obtained via GPS signals and Network Time Protocol software, as described in \citet{2004ApJ...605..846N}.  These times were manually cross-checked  nightly against the official observatory clock and a digital wristwatch synced daily with the WWVB time signal operated by the National Institute of Standards and Technology; the clocks were fully consistent at all times.  Table \ref{tab.obslog} gives the  basic parameters of each individual target run. 

We reduced the data using the pipeline and methods of \citet{2005ApJ...625..966M,2008ApJ...676..573M}.  In summary, we performed weighted aperture photometry for a wide variety of aperture radii; we compared the noise in the discrete Fourier transform (DFT)  for each choice of aperture; and we selected the single aperture radius with the lowest noise in the DFT.  We divided the resulting light curve by a weighted combination of one to three neighboring stars as similar in color to the targets as possible.  We removed differential atmospheric extinction by dividing the light curve with the best fitting  second-order polynomial.  We excluded portions of the light curve from the analysis that are clearly impacted by events such as cosmic ray hits, thick clouds, and highly volatile seeing.  We accounted for all UTC leapseconds and corrected exposure midtimes to Barycentric Dynamical Time (TDB) using the methodology of \citet{1980A&AS...41....1S}.

In order to reduce the noise in the DFT and search for low-amplitude signals, we combined multiple runs for the same object.  This technique is only valid if the modulations remain coherent between the runs.  When possible, we bootstrap so that we can determine a period precise enough to bridge the gap between runs without losing track of the cycle count, as described  in, e.g., \citet{1985ApJ...292..606W}.  Due to issues with weather and scheduling, we were not able to fully bootstrap the three previously known variable observations.  Since we had found a frequency that was consistent with coherence on this timescale, we proceeded with the analysis under the assumption that the observed modulations were stable (see discussions in Section \ref{sec.knowndqvs}).

\section{Previously-known variables}\label{sec.knowndqvs}

\begin{figure}
\begin{center}
\includegraphics[height=0.99\columnwidth,angle=270]{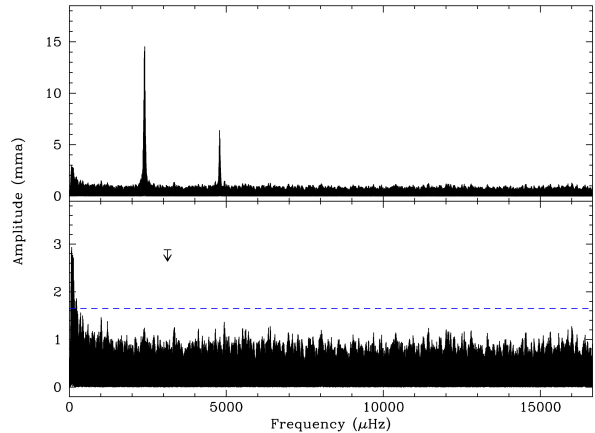}
\end{center}
\caption{Discrete Fourier transforms (DFTs) for the combined observations of SDSS J1426+5752.  The top panel shows the DFT of the reduced light curve; the bottom panel shows the transform after prewhitening the observations by the best-fit fundamental and harmonic.  The horizontal dashed line indicates the amplitude above which a peak would have a false alarm probability of $\leq 1\%$.  The upper limit symbol indicates the frequency and amplitude of the claimed 319.7 s periodicity of \citet{2009ApJ...702.1593G}, which is not detected in our data.  Outside of the fundamental and harmonic, no significant signals are observed. \label{fig.1426dft}}
\end{figure}

\begin{figure}
\begin{center}
\includegraphics[height=0.99\columnwidth,angle=270]{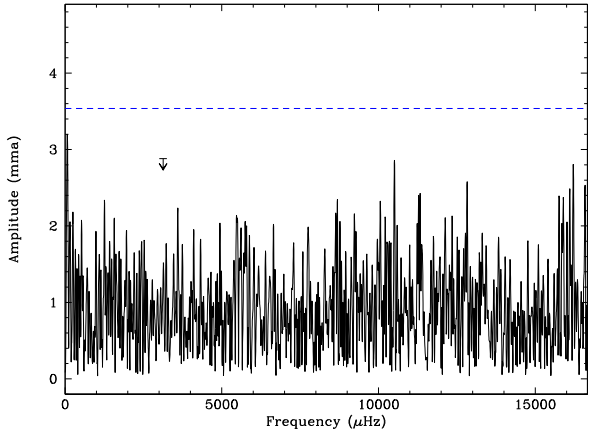}
\end{center}
\caption{DFT for run A1712 on SDSS J1426+5752 after prewhitening by the two harmonically related periods.  This represents our sole run overlapping with the \citet{2009ApJ...702.1593G} data.  As with the combined data set, no evidence is observed for a mode with a period of 319.7 s; the claimed amplitude and frequency from \citet{2009ApJ...702.1593G} is marked with the upper limit.  For these data, the 1\% false alarm probability amplitude threshold is 3.5 mma and is indicated by the horizontal dashed line. \label{fig.a1712}}
\end{figure}

\subsection{SDSS 1426+5752}\label{sec.1426}
SDSS 1426+5752 is the prototype DQV.  In the discovery paper, we reported detections of modulations with periods of 417.66 s and 208.82 s with amplitudes of 17 mma and 7 mma, respectively (1 mma refers to a fractional amplitude of $10^{-3}$, or 0.1\%).  We also claimed a tentative detection of modulation near the harmonically-related period of 83.5 s \citep{2008ApJ...678L..51M}.    Follow-up observations by \citet{2009ApJ...702.1593G} refined the periods and amplitudes of the two larger-amplitude modes, but refuted the presence of the 83.5 s periodicity.  They also claim a significant detection of non-harmonically related modulation with a period of $\approx 319.7$ s and an amplitude of 2.88 mma.  

Our observations span just over three years with a total exposure time of nearly 270 ks.  Figure \ref{fig.1426dft} shows the DFT of the combined observations.  The two primary periods of modulation are clearly visible.  After prewhitening the data by these two modulations, we see no other significant amplitude peaks (Figure \ref{fig.1426dft}, bottom panel). 
We determine the false alarm probabilities using Equation 18 of \citet{1982ApJ...263..835S}.  The horizontal dashed line in Figure \ref{fig.1426dft} indicates the $P=0.01$ false alarm level; i.e., any amplitude peaks above this level have a $\leq 1\%$ chance of being false detections.  We do not consider the apparent signal at low frequencies to be convincing due to $1/f$ noise and to atmospheric phenomena described in Section \ref{sec.0005}.

In particular, we see no evidence of a 3 mma signal with the 319 s period discussed by \citet{2009ApJ...702.1593G}, either in the combined data set or in individual runs.  This implies that the 319 s period is either not coherent, significantly variable in amplitude, or was a spurious result.  We have a  single time-series run from one night overlapping with the \citet{2009ApJ...702.1593G} data, A1712.  We show the DFT from that single time-series observation in Figure \ref{fig.a1712} after pre-whitening by the overall best-fit periods; we detect no evidence for the 319 s period in this run.  However, the \citet{2009ApJ...702.1593G} data cover a time frame nearly one month prior to our run, and it is possible that the mode's amplitude weakened prior to our observation.  In short, we are unable confirm if this period were present at the time of their observations.

\begin{figure}
\begin{center}
\includegraphics[width=0.99\columnwidth]{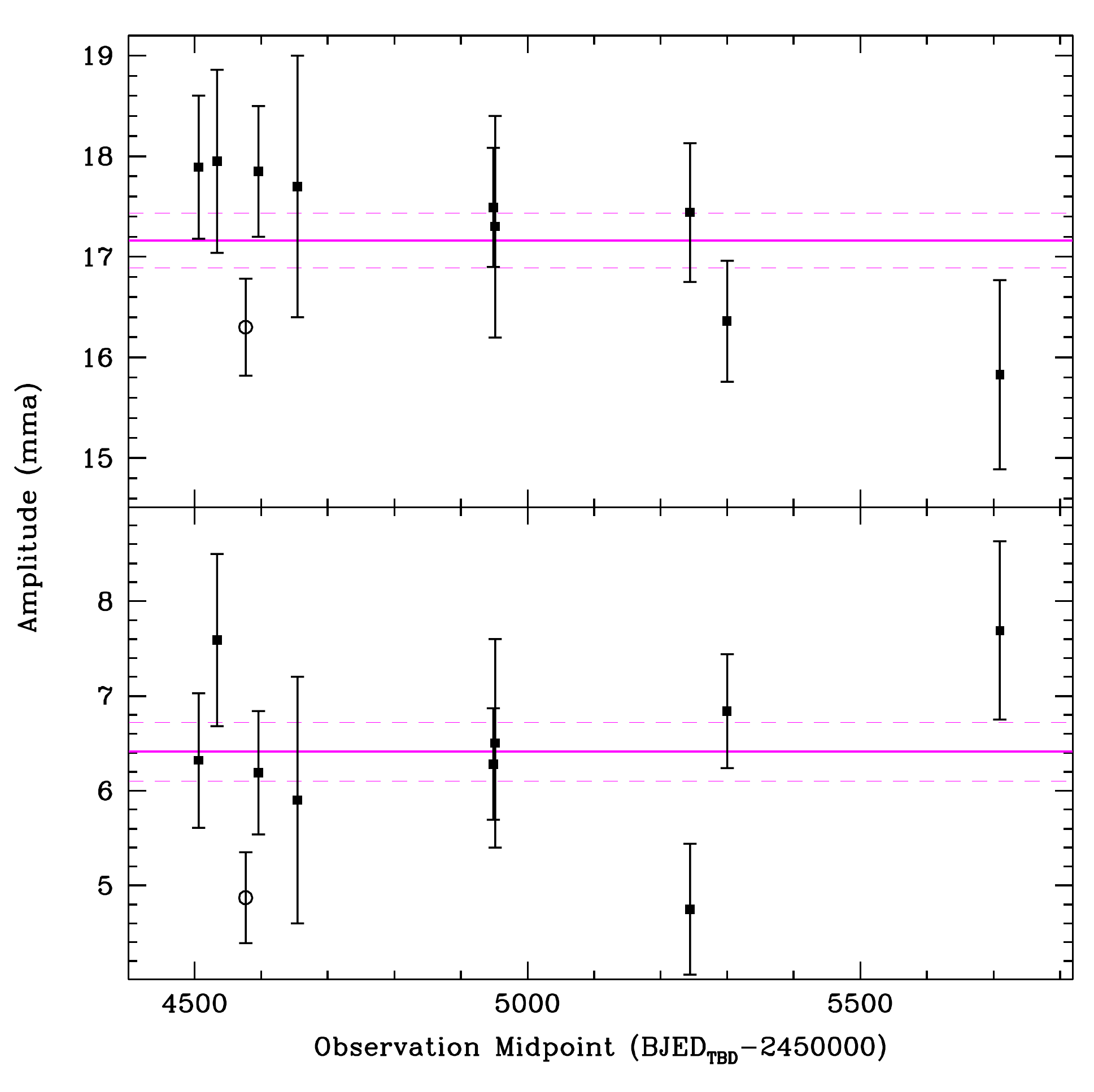}
\end{center}
\caption{Combined monthly amplitudes for modulations in SDSS J1426+5752.  The top panel contains amplitudes of the fundamental 417.7 s modulation; the lower panel shows the harmonic 208.9 s modulation amplitudes.  Filled squares with error bars are the data from this paper; the open circles are measurements from   \citet{2009ApJ...702.1593G}.  Solid magenta lines are the measured amplitudes from the combined set of all data; dashed magenta lines are the $1\sigma$ uncertainties.  The fundamental modulation may show a slow decrease over time, but all points are consistent with the overall best-fit amplitude.  There is no evidence for significant variability in the harmonic modulation. \label{fig.1426_amps}}
\end{figure}

The best-fit periods for the modulations observed in SDSS J1426+5752 are 417.706776(14) s and 208.853366(11) s; within the stated errors, the shorter period is exactly half of the fundamental period, greatly strengthening the hypothesis that these periods are harmonically related.  The amplitudes and phases of these modulations are given in Table \ref{tab.1426}.  We also calculate amplitudes and phases for monthly subsets of our data using the best overall frequency in order to check for variations in amplitude or phase.  As seen in Figure \ref{fig.1426_amps}, there is a suggestion that the amplitude of the 417 s modulation may be decreasing with time.  However, we have no additional data at this time with which to confirm this trend.  The harmonically-related 208.9 s modulation shows no evidence of significant amplitude variations, and both modulations' phases are consistent with being constant.

\setcounter{table}{1}
\begin{deluxetable}{lcccc}
\tablecolumns{5}
\tablewidth{0pt}
\tablecaption{Periods, amplitudes, and phases for modes observed in SDSS J1426+5752 \label{tab.1426}}
\tablehead{\colhead{Run} & \multicolumn{2}{l}{$P_1=417.706776(14)$ s} & \multicolumn{2}{l}{$P_2=208.853366(11)$ s} \\  \colhead{Subset} & \colhead{$A_1$ (mma)} & \colhead{$T_{0,1}$ (s)} & \colhead{$A_2$ (mma)} & \colhead{$T_{0,2}$ (s)} }
\startdata
All &  $17.16 \pm 0.27$            & $91.3\pm 1.8$ & $6.41\pm 0.31$ & $30.6\pm 2.7$ \\
2008 Feb & $17.89\pm 0.71$ & $90.2\pm 2.6$ & $6.32\pm 0.71$ & $28.9\pm 3.7$ \\
2008 Mar & $17.95\pm 0.91$ & $92.6\pm 2.6$ & $7.59\pm 0.91$ & $28.6\pm 4.0$ \\
2008 May & $17.85\pm 0.65$ & $95.6\pm 2.6$ & $6.19\pm 0.65$ & $32.5\pm 3.5$ \\
2008 Jul & $17.7\pm 1.3$        & $90.8\pm 2.6$ & $5.9\pm 1.3$     & $32.6\pm 7.1$ \\
2009 Apr & $17.49\pm 0.59$  & $90.7\pm 2.6$ & $6.28\pm 0.59$ & $32.7\pm 3.1$ \\
2009 May & $17.3\pm 1.1$     & $91.3\pm 2.6$ & $6.5\pm 1.1$      & $37.5\pm 5.6$ \\
2010 Feb & $17.44\pm 0.69$ & $93.2\pm 2.6$ & $4.75\pm 0.69$ & $32.1\pm 4.8$ \\
2010 Apr & $16.36\pm 0.60$ & $84.4\pm 2.6$ & $6.84\pm 0.60$  & $25.5\pm 2.9$ \\
2010 May & $15.83\pm 0.94$ & $100.5\pm 2.6$ & $7.69\pm 0.94$ & $33.6\pm 4.0$ \\
\enddata
\tablecomments{$T_0$ is the start of the first cycle, in seconds after BJD$_{\rm TDB}=2454506$ d. }
\end{deluxetable}

\subsection{SDSS J2200$-$0741}\label{sec.2200}
SDSS J2200$-$0741 was recognized as a variable by \citet{2008ApJ...688L..95B}, with follow-up observations reported by \citet{2009ApJ...703..240D}.   SDSS J2200$-$0741 is unique among the variable hot DQs as the harmonic modulation ($P=327.2$ s) has an amplitude nearly as high as the fundamental ($P=654.4$ s).    \citet{2009ApJ...703..240D} also claim detection of two harmonically unrelated periods ($P=577.6$ s and $P=254.7$ s) at amplitudes of $\approx 1.5$ mma and the second harmonic of the fundamental ($P=218.1$ s) with an amplitude of $\approx 1$ mma.

\begin{figure}
\begin{center}
\includegraphics[height=0.99\columnwidth,angle=270]{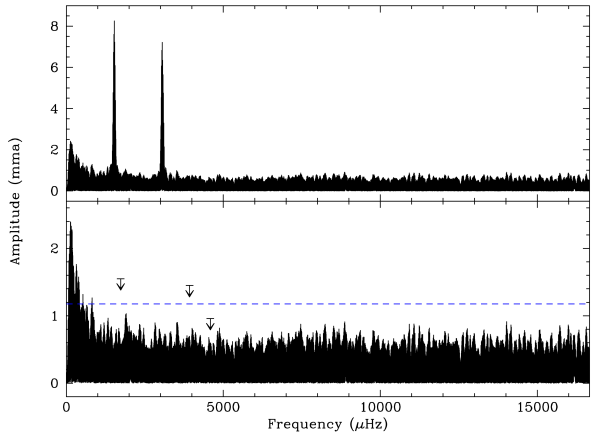}
\end{center}
\caption{DFTs for the combined observations of SDSS J2200$-$0741.  The top panel contains the DFT of the extracted light curve; the bottom panel shows the DFT after prewhitening the data by the best-fit fundamental and harmonic.  The horizontal dashed line indicates the amplitude threshold above which a peak would have a false alarm probability of $\leq 1\%$.  The upper limit symbols indicate the frequencies and amplitudes of additional periodicities claimed by \citet{2009ApJ...703..240D}, which are not present in these data.  Outside of the fundamental and harmonic, no significant signals are observed. \label{fig.2200dft}}
\end{figure}

Our data for SDSS J2200$-$0741 total 162 ks obtained over two years.  Figure \ref{fig.2200dft} shows the DFT of our combined data set (top) and the combined data set after prewhitening by the two primary periodicities (bottom).  We find no evidence of modulations at frequencies other than the fundamental and harmonic.  As our observations were obtained more than seven weeks after those of \citet{2009ApJ...703..240D}, we cannot rule out that changing amplitudes resulted in the non-harmonic modes they observed being undetectable in our data.

We find that the best fitting period for the fundamental modulation in SDSS J2200$-$0741 is $P_1=654.427851(74)$ s with an amplitude of $8.31\pm 0.18$ mma, while the apparent harmonic has a period of $P_2=327.213999(21)$ s with an amplitude of $7.40\pm 0.18$ mma.  $2P_2$ is mildly inconsistent with $P_1$ at the $\approx 2\sigma$ level, leaving the question of a precise harmonic relationship open, though the difference is at a fractional level of $2\times 10^{-7}$.

Table \ref{tab.2200} gives the amplitudes and phases of the two significant modulations, obtained by holding the periods fixed at the best-fit values.  Wth the exception of the fundamental modulation's phase determination for the 2009 Sep data, the amplitudes and phase are consistent with being constant.  The reason for this deviation is unclear; conditions for both nights were acceptable, with variable thin cirrus, light winds, and steady 1\farcs 4 seeing, so it is possibly intrinsic to the star.

\begin{deluxetable}{lcccc}
\tablecolumns{5}
\tablewidth{0pt}
\tabletypesize{\small}
\tablecaption{Amplitudes and phases for SDSS J2200$-$0741 \label{tab.2200}}
\tablehead{\colhead{Run} & \multicolumn{2}{l}{$P_1=654.427851(74)$ s} & \multicolumn{2}{l}{$P_2=327.213999(21)$ s} \\ \colhead{Subset} & \colhead{$A_1$ (mma)} & \colhead{$T_{0,1}$ (s)} & \colhead{$A_2$ (mma)} & \colhead{$T_{0,2}$ (s)} }
\startdata
All & $ 8.31\pm 0.18$ & $37.4\pm 4.6$ & $7.40\pm 0.18$ & $268.1\pm 2.6$ \\
2008 Oct & $7.97\pm 0.51$ & $44.3\pm 6.6$ & $7.23\pm 0.51$ & $265.2\pm 3.7$ \\
2009 Sep & $7.4 \pm 1.3 $ & $-15\pm 18$ & $8.5\pm 1.3$ & $275.0\pm 8.0$ \\
2009 Nov & $8.23\pm 0.67$ & $41.9\pm 8.4$ & $6.69\pm 0.67$ & $271.6\pm 5.2$\\
2010 Sep & $8.19\pm 0.57$ & $50.8\pm 7.3$ & $7.61\pm 0.57$ & $264.3\pm 3.9$\\
2010 Oct & $9.13\pm 0.36$ & $35.5\pm 4.1$ & $7.33\pm 0.36$ & $267.8\pm 2.6$\\
\enddata
\tablecomments{$T_0$ is the start of the first cycle, in seconds after BJD$_{\rm TDB}=2454506$ d.}
\end{deluxetable}

\subsection{SDSS J2348$-$0943}\label{sec.2348}
SDSS J2348$-$0943 was discovered to be variable by \citet{2008ApJ...688L..95B}, who detected a 1052 s modulation with an amplitude of $\sim 7$ mma and a formally significant detection of a non-harmonically related modulation of 379 s, though they discuss that this second modulation is likely spurious.  \citet{2009ApJ...703..240D} confirm an 8 mma modulation with a period of 1044.2 s, do not detect any modulation at 379 s, and claim a significant detection of a 3.6 mma non-harmonically related periodicity at 416.9 s.  Neither group detects any harmonically related periodicities.

\begin{figure}
\begin{center}
\includegraphics[height=0.99\columnwidth,angle=270]{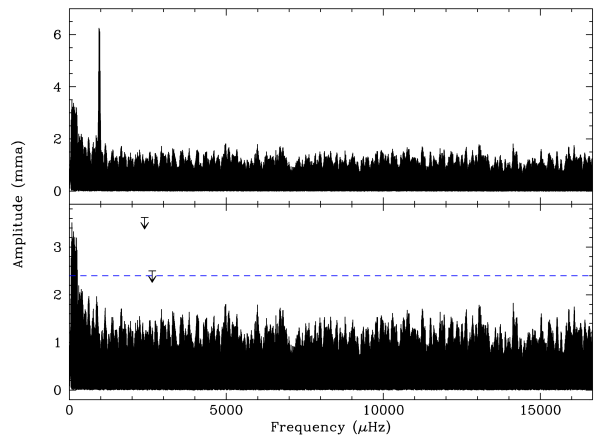}
\end{center}
\caption{DFTs for the combined observations of SDSS J2348$-$0943.  The top panel displays the DFT of the reduced light curve; the bottom panel shows the DFT after prewhitening the data by the best-fit modulation.  The horizontal dashed line indicates the amplitude above which a peak would have a false alarm probability of $\leq 1\%$.  The upper limit symbols indicate the frequencies and amplitudes of a potential 379 s periodicity proposed by  \cite{2008ApJ...688L..95B} and a 416.9 s periodicity tentatively claimed by \citet{2009ApJ...703..240D}, neither of which are detected in these data.\label{fig.2348dft}}
\end{figure}

Our observations total 164 ks over a two-year baseline; we detect only one significant modulation over this time; see Figure \ref{fig.2348dft}.  We are able to constrain the period of the modulation to $P=1044.22105(42)$ with a mean amplitude of $6.82\pm 0.38$ mma.  As with \citet{2009ApJ...703..240D}, we do not detect any harmonically-related modulations, and we do not detect the 379 s period mentioned by \citet{2008ApJ...688L..95B}.  We also do not detect any modulation with a 416.9 s period at the amplidtude claimed by \citet{2009ApJ...703..240D}.  The amplitudes and phases of the modulations calculated from monthly subsets of data are consistent with being constant, as shown in Table \ref{tab.2348}.

\begin{deluxetable}{lcc}
\tablecolumns{3}
\tablewidth{0pt}
\tabletypesize{\small}
\tablecaption{Amplitudes and phases for SDSS J2348$-$0942 \label{tab.2348}}
\tablehead{\colhead{Run} & \multicolumn{2}{l}{$P=1044.22105(42)$ s}  \\ 
\colhead{Subset} & \colhead{$A$ (mma)} & \colhead{$T_{0}$ (s)} }
\startdata
All & $6.82\pm 0.38$ & $261\pm 13$ \\
2008 Oct & $6.58\pm 0.46$ & $257\pm 12$ \\
2008 Dec & $7.0\pm 1.7$ & $250\pm 40$ \\
2009 Oct & $6.58\pm 0.96$ & $285\pm 24$ \\
2009 Nov & $9.9\pm 1.7$ & $265\pm 29$ \\
2010 Oct & $6.06\pm 0.78$ & $260\pm 21$ \\
\enddata
\tablecomments{$T_0$ is the start of the first cycle, in seconds, after BJD$_{\rm TDB}=2454506$ d. }
\end{deluxetable}

\begin{figure}
\begin{center}
\includegraphics[width=0.99\columnwidth]{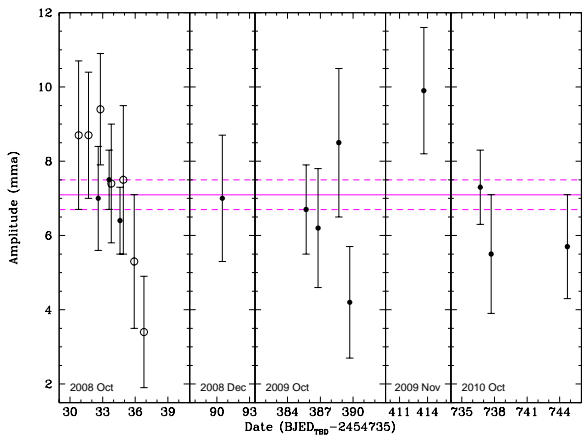}
\end{center}
\caption{Measured nightly amplitudes of the $P=1044.2$ s modulation of SDSS J2348$-$0942.  Open circles are data from \citet{2009ApJ...703..240D}, while filled circles are data from this paper.  The solid magenta line is the best-fit amplitude of 6.82 mma; flanking dotted lines are the $1\sigma$ uncertainties. The horizontal axis is chosen to match that used in  \citet{2009ApJ...703..240D}.  Where the data overlap, both data sets show consistent amplitudes.  Our data are fully consistent with no variations in amplitude. \label{fig.2348_amps}}
\end{figure}

\citet{2009ApJ...703..240D} discuss a potential modulation in the amplitude on a timescale of days.  As our 2008 Oct data overlap in time, in Figure \ref{fig.2348_amps} we present the best-fit amplitudes for each of our nightly object runs over the entire two years of data, as well as those from \citet{2009ApJ...703..240D}.  We note that, on the few nights that overlap, both studies' amplitudes are consistent, despite slight differences in filter and instrument responses.  Unfortunately, we were unable to obtain data on the nights where \citet{2009ApJ...703..240D} claim the best evidence for amplitude modulation, and in later months we were simply unable to string together multiple nights, primarily due to inclement weather.

\subsection{Summary of Previously Known Variables}
We have obtained time-series photometry spanning multiple years for three of the previously known variable hot DQ WDs.  We are able to confirm the primary modulation in all three hot DQs, as well as the first harmonic in SDSS J1426+5752 and SDSS J2200$-$0741.  These modulations appear to be coherent, allowing us to combine all our data and determine precise periods, amplitudes, and phases.  Each of the hot DQs shows signs of potential amplitude variations in the primary modulation, but the data are not statistically significant in any one case.

However, we are unable to confirm any of the non-harmonically related modulations mentioned in other publications; our upper limits on detections are robust in most cases.  Due to a paucity of overlapping data, we cannot rule out amplitude variations as the reason behind our nondetections.  Perhaps non-harmonically related modulations are simply very short-lived in these stars, or perhaps these detections were statistical flukes.

\section{Hot DQs not observed to vary}

\begin{figure}
\begin{center}
\includegraphics[width=0.99\columnwidth]{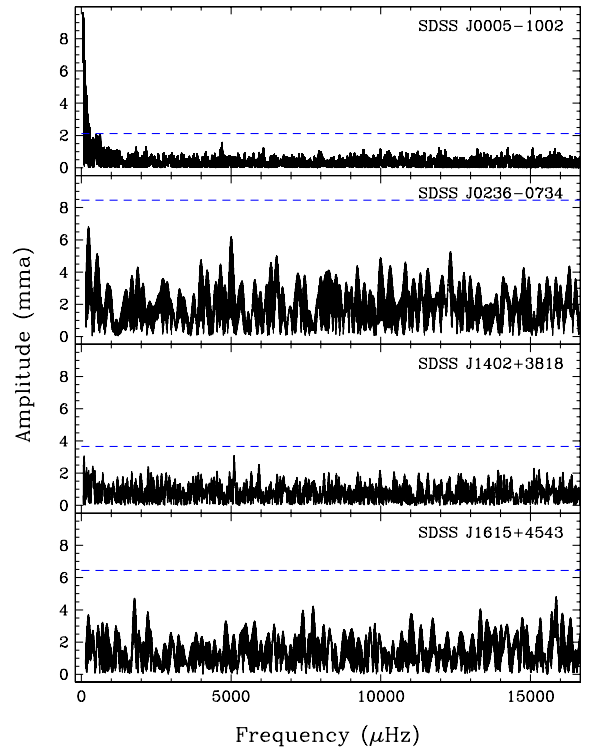}
\end{center}
\caption{DFTs of four hot DQs not observed to vary with short periods within our detection limits.  Horizontal dashed lines indicate the 1\% false alarm probability threshold for the given observations.  SDSS J0005$-$1002 is known to vary with a period of 2.1 d; the excess power at very low frequencies is likely indicative of that variability.  Only SDSS J0005$-$1002 has a signal-to-noise level sufficient to detect the smallest known amplitudes in DQVs  ($\sim 3$ mma). \label{fig.novs}}
\end{figure}

In addition to the three known variable hot DQ WDs, we present observations of four hot DQs in which we fail to find short-period variability ($P\lesssim 1$ h).  Figure \ref{fig.novs} shows the DFTs for the combined light curves of each object.  Our detection limits are indicated in the figure by horizontal dashed lines and are given in Table \ref{tab.novs}; these represent the 1\% false alarm probability thresholds.  

\begin{deluxetable}{lc}
\tablecolumns{2}
\tablewidth{0pt}
\tablecaption{Upper limits on undetected mode amplitudes \label{tab.novs}}
\tablehead{\colhead{Object} & \colhead {1\% FAP (mma)}}
\startdata
SDSS J0005$-$1002 & 2.1 \\
SDSS J0236$-$0734 & 8.5 \\
SDSS J1402+3818 & 3.7 \\
SDSS J1615+4543 & 6.4 \\[10 pt]
SDSS J1426+5752 & 1.6 \\
SDSS J2200$-$0751 & 1.2 \\
SDSS J2348$-$0942 & 2.4 \\
\enddata
\tablecomments{The table gives the lower limits for mode amplitudes that would have a $\leq 1\%$ false alarm probability.  SDSS J1426+5752, SDSS J2200$-$0741, and SDSS J2348$-$0942 limits are calculated after prewhitening the combined data by the modes identified in Table \ref{tab.1426}, Table \ref{tab.2200}, and Table \ref{tab.2348}, respectively.  The calculation for SDSS J0005$-$1002 ignores frequencies below 300 $\mu$Hz for reasons described in the text.}
\end{deluxetable}

The lowest-amplitude modulations observed in any hot DQs are those in SDSS J1337$-$0026\footnote{Long format: \objectname[SDSS J133710.19-002643.8]{SDSS J133710.19$-$002643.8}}, which has harmonically-related modulations with amplitudes of 2.6 and 3.3 mma \citep{2010ApJ...720L.159D}.  In order for a non-detection to be meaningful in the context of hot DQs, our 1\% false alarm probability thresholds need to exceed these small levels.  Under these criteria, our non-detections of periodic modulations in SDSS J0236$-$0734 and SDSS J1615+4543 (1\% FAP at 8.5 and 6.4 mma, respectively) are not stringent limits on variability in these stars.  However, our non-detection of variability in SDSS J1402+3818, with a 1\% false alarm probability threshold of  3.7 mma, approaches this criterion.  If the false alarm probability level can be pushed $\sim 1$ mma fainter, SDSS J1402+3818 would be the only hot DQ WD \emph{not} to exhibit variability at the 3 mma level.

\subsection{SDSS J0005$-$1002}\label{sec.0005}

Much more interesting than our three hot DQs not observed to vary is our non-detection of short-period variability in SDSS J0005$-$1002.  \citet{2013MNRAS.433.1599L} find that this hot DQ is variable, but with a large amplitude of 11\% and a 2.11 d period.  They interperet this variability as rotation for multiple reasons: there is no known mechanism to maintain standing waves in nonradial pulsations of such a long period, many magnetic WDs are observed to rotate on timescales of hours to days through both photometric and spectropolarimetric variations, and many non-DQ pulsating WDs show splitting indicative of rotation with periods of hours to days.  

Prior to the discovery of rotational variability in this hot DQ by \citet{2013MNRAS.433.1599L}, we had noticed excessive power at the lowest frequencies in the DFT of the star (see the top panel of Figure \ref{fig.novs}).  However, we postulated that this power was due to a combination of the spectral energy distribution of hot DQs and atmospheric phenomena common at McDonald Observatory.  Due to strong line blanketing in the ultraviolet, flux is redistributed to short optical wavelengths \citep{2005ASPC..334..253B,2008ApJ...683..978D}.  This results in a spectrum that is steeper than the Rayleigh-Jeans tail of a blackbody spectrum in the broad bandpass defined by our filter.  This results in none of the comparison stars having  optical colors as blue as the hot DQs we are monitoring.  Therefore, atmospheric phenomena that selectively affect the transmission of short wavelengths will not be corrected in our reductions.

While observing hot DQs, we noted multiple phenomena that clearly introduced artifacts into our time-series data.  We noticed preferrential dimming of hot DQs on nights when shifting winds would bring soot (and a strong odor of burnt mesquite) from distant wildfires toward the observatory.   Many nights a wandering atmospheric dry line would pass through the observatory once or twice, causing sudden shifts in temperature, humidity, and wind direction as well as short periods of greatly worsened seeing; these passages also preferentially affected the hot DQ photometry.  The 2009 Apr 30 observations of SDSS J1426+5752 exhibited significant changes in atmospheric transparancy from an unknown source that again affected the hot DQ more than comparison stars.    The 2nd-order polynomial fit applied to all of our light curves to remove differential atmospheric extinction also would greatly weaken any true signal with $P\gtrsim 2$ h in our data.  For these reasons, we ignored the low-frequency signal in SDSS J0005$-$1002, most of which was likely due to true photometric variations.  

\section{Rotation: A Single Source of Variability in all Hot DQs?}
Several studies have now presented observations of hot DQs and their progeny, and clear trends are emerging that are strengthened by our data.  We exclude the hot DQ WDs not observed to vary from this analysis, SDSS J0236$-$0734, SDSS J1402+3818, and SDSS J1615+4543, due to the insufficient signal-to-noise in their DFTs.  We include SDSS J1036+6522; although it does not exhibit \ion{C}{2} like true hot DQs, our analysis strongly suggests that it is evolutionarily related to the hot DQs \citep{2013ApJ...769..123W}.  

From the collected data, we emphasize two seemingly universal traits of variability in hot DQs:
\begin{itemize}
\item \emph{All} hot DQ WDs show significant, periodic modulations.  These modulations are  coherent over the observational baselines.
\item With the possible exception of SDSS J1153+0056\footnote{Long format name: \object{SDSS J115305.55+005646.2}} \citep{2011ApJ...733L..19D}, \emph{all} hot DQ WDs exhibit only harmonically-related modulations.  Non-harmonic modulations have been claimed, but our non-detection of these means either these modulations are not coherent over long times scales or they are spurious.  Given this, and given that the time-series observations of SDSS J1153+0056 are limited to a single 2087 s HST observation, we are skeptical of the claims of non-harmonically related modes in this WD.
\end{itemize}

These two traits, along with the detailed properties of each star and of the observed modulations, are difficult to reconcile with the hypothesis that the variability is caused by nonradial pulsations in the stars.  SDSS J1036+6522 is much cooler than the various predicted instability strips \citep{2009A&A...506..835C,2008A&A...483L...1F,2008ApJ...678L..51M}, and the period of modulations in SDSS J0005$-$1002 is too long for coherent pulsations.  The pulse shapes of the variables wih harmonically-related periods are very different from the pulse shapes of known pulsators in other spectral classes of WDs \citep{2008ApJ...678L..51M,2008ApJ...688L..95B,2009ApJ...703..240D}. Most of the variable hot DQs in general have strong magnetic fields ($\sim 1$ MG), while no strongly magnetic pulsators exist in ofther spectral classes \citep[e.g.,][]{2008ApJ...683L.167D}.  Although some pulsating WDs exhibit only a single independent mode, most pulsate at multiple non-harmonically related frequencies. 

While there may be mechanisms for reconciling these issues with nonradial pulsations, there may be a simpler alternative: rotation. Rotation of highly magnetic SDSS J0005$-$1002 is the most plausible explanation for its variability; this is discussed in convincing detail by \citet{2013MNRAS.433.1599L}, who also suggest rotation as the cause of variability in other hot DQs.  If we conservatively assume that the longest observed period in each variable hot DQ is the rotation period, then most hot DQs are among the fastest known WD rotators.  

Is rapid rotation consistent with proposed formation scenarios for hot DQs?  We \citep{2006ApJ...643L.127W,2013ApJ...769..123W} and \citet{2010Sci...327..188G} have previously discussed the possibility that hot DQs and cooler oxygen-rich atmosphere WDs are the remnants of the super-asymptotic giant branch stars; WD rotation periods as low as $\approx 250$ s are possible from such stars \citep{2004IAUS..215..561K,2015ASPC..493...65K,2013ApJ...775L...1T}.  Variable core-envelope coupling strengths or magnetic brakng could explain the spread in observed periods.

Based on kinematic arguments, \citet{Dunlap2015} and \citet{2015ASPC..493..547D} propose that hot DQs are the remnants of recent mergers of double degenerate systems insufficiently massive to spawn Type Ia supernovae.  Such mergers have been invoked to explain the presence of magnetic fields in massive WDs \citep[e.g.,][]{2005MNRAS.356..615F,2012ApJ...749...25G}.  Rapid rotation would be a natural result of transfer of orbital angular momentum to the merger remnants, and \citet{2015ASPC..493..547D} interpret the variability in hot DQs as this rotation.

\citet{2007Natur.450..522D} suggest that hot DQ WDs may be the result of a vigorous late thermal pulse creating a carbon-rich PG1159 star such as \object{H1504+65}.  Gravitational settling of carbon and oxygen followed by renewed dredge-up by deepening convective zones leads to an evolving spectral type for these stars.  Calculations by \citet{2009A&A...494.1021A,2009ApJ...693L..23A} show that this may be a viable scenario, especially for the high surface gravity.  \cite{2009ApJ...693L..23A} note that the presence of magnetism has yet to be considered; presumably rapid rotation also affects models of hot DQ evolution and spectral characteristics.  However, rapid rotation does not appear to be a necessary ingredient in the current models.

One observation of the variability in hot DQs that may redeem the pulsational model is that of the ultraviolet (UV) -to-optical ratio of the amplitudes in fractional flux.  \citet{2011ApJ...733L..19D} find this ratio to be $\sim 2-4$ from HST UVobservations of variable hot DQs.  This ratio is comparable to that observed in known nonradial pulsators \citep[e.g.,][]{1985ApJ...289..774H,1994AJ....107..298K} and that predicted from model atmospheres of DA WDs \citep[e.g.,][]{Fontaine1996}.  The observed UV-to-optical fractional flux amplitude ratio may also be consistent with a cool spot at one of the magnetic poles of a variable hot DQ, as cooler temperatures would suppress UV flux more than optical.  A detailed analysis of a spot model calculating the fractional flux amplitude ratio is necessary, however, as the UV spectra of hot DQs is heavily line blanketed and horribly complex \citep{2011ApJ...733L..19D}.  

Occam's Razor would favor rotation as the source of variability in all hot DQ stars, as rotation can neatly explain most of the observations of variability in hot DQ WDs. However, rotation needs to be proven, especially for the short-period variables.  Back-of-the-envelope calculations suggest that phase-resolved spectropolarimetry should be feasible for at least one or two short-period variable hot DQs with existing facilities.  We \emph{cannot} rule out non-radial pulsations as a viable source of variability, as the basic arguments that led us to suggest non-radial pulsations still stand.  However, pulsational mechanisms have an increasing list of oddities in the hot DQ variability properties as compared with known classes of pulsating WDs that needs to be explained in order to remain a viable mechanism.

In the near term, we suggest that the following observations and theoretical modeling may be the most fruitful means of finally cracking the tough case of the mysteriously variable hot DQs:
\begin{enumerate}
\item Additional optical observations of hot DQs not observed to vary, especially SDSS J1402+3818, in order to push down noise in the search for low-amplitude and/or long-period modulations.
\item Searches for long-term variabilty in known variable hot DQs, in order to rule out more typical WD rotation periods of hours to days.
\item Polarimetric and or spectropolarimetric observations of one or more short-period variable hot DQs in order to look for evidence of rotation harmonically related to the periods of the observed modulations.
\item Searches for variability in the second-sequence cool DQ WDs thought to be the later stage of hot DQ evolution in order to determine the evolution of the modulation periods.
\item More detailed evolutionary models for each of the proposed origins of hot DQs that include the presence of MG-level magnetic fields and rapid rotation in order to determine which progenitor scenarios are indeed viable.
\end{enumerate}

\acknowledgements 
We would like to thank several members of our collaboration for their assistance in obtaining these data: D.~Chandler, S.~DeGennaro, J.~J.~Hermes, A.~Mukadam, W.~Strickland, and B.~Walter.   In addition, J.~Kuehne and R.~Leck led long, difficult, but ultimately successful efforts to upgrade the Struve Telescope and Argos data collection software during the course of this project, increasing data quality and productivity.  We mourn the loss of Ed Nather, who was an integral part of our team for decades, and celebrate the legacy he leaves behind in high-speed photometry at McDonald Observatory.  We thank P.~Dufour, G.~Fontaine, and B.~Dunlap for multiple in-depth discussions concerning hot DQs throughout this project.  This work was done in part through the REU Program in Physics and Astronomy at Texas A\&M University-Commerce funded by the National Science Foundation under grant PHY-1359409. KAW also acknowledges the support of NSF award AST-0602288 and support for this work from the GALEX Guest Investigator program under NASA grant NNX11AG82G during the data collection for this project.  R.E.F, M.H.M., and D.E.W. gratefully acknowledge the support of the NSF under grant AST-1312983.   This research has made use of NASA's Astrophysics Data System and of the SIMBAD database, operated at CDS, Strasbourg, France.

{\it Facility:} \facility{Struve (Argos)}


\clearpage
\setcounter{table}{0}
\begin{turnpage}
\begin{deluxetable*}{lccccccc}
\tablecolumns{8}
\tablewidth{0pt}
\tabletypesize{\scriptsize}
\tablecaption{Observing log\label{tab.obslog}}
\tablehead{\colhead{Object\tablenotemark{a}} & \colhead{$g$\tablenotemark{a}} & \colhead{$u-g$\tablenotemark{a}} & \colhead{Run ID} & \colhead{UT Date} & \colhead{Start\tablenotemark{b}} & \colhead{Exp.~Time} & \colhead{Duration} \\ & (mag)  & (mag)  & & & (BJD$_{\rm TDB}$-2450000) & (s) & (s) \\ }
\startdata
\objectname[SDSS J000555.91-100213.4]{SDSS J000555.91$-$100213.4} & $17.71\pm 0.02$ & 
	$-0.38\pm 0.03$  & A1986 & 2009 Sept 16 & 5090.73375732 & 10 &  22200 \\ 
(SDSS J0005$-$1002) & & & A2014 & 2009 Oct 23 & 5127.76784051 & 15 & 7380 \\ 
	& & & A2018 & 2009 Oct 24 & 5128.71185114 & 15 & 10815  \\ [3pt] 
 \objectname[SDSS J023637.42-073429.5]{SDSS J023637.42$-$073429.5} & $19.77\pm 0.05$ & 
 	$-0.50\pm 0.06$ & A1655 & 2008 Feb 09 & 4505.56853031 & 30 & 7050   \\ 
(SDSS J0236$-$0734) & & & A1659 & 2008 Feb 10 & 4506.57706195 & 30 & 6060 \\ [3pt] 
\object{SDSS J140222.25+381848.9} & $18.99\pm 0.02$ &
  	$-0.30\pm 0.04$ & A1716 & 2008 Jul 05 & 4652.62357901 & 30 & 16320 \\ 
(SDSS J1402+3818)	& & & A1717 & 2008 July 06 & 4653.62484307 & 30 & 15600 \\ [3pt] 
\object{SDSS J142625.71+575218.4	} & $19.14\pm 0.02$ & 
	$-0.37\pm 0.04$ & A1661 & 2008 Feb 10 & 4506.82619867 & 30 & 18000  \\ 
(SDSS J1426+5752)	& & & A1663 & 2008 Feb 11 & 4507.82970617 & 30 & 10020 \\ 
	&&& A1666 & 2008 Mar 08 & 4533.76152116 & 15 & 22500 \\ 
	&&& A1712 & 2008 May 09 & 4595.65322666 & 30 & 26460 \\ 
	&&& A1719 & 2008 Jul 07 & 4654.62145901 & 30 & 10020 \\ 
	&&& A1864 & 2009 Apr 24 & 4945.86516430 & 30 & 9900 \\ 
	&&& A1867 & 2009 Apr 27 & 4948.60621111 & 30 & 31590 \\ 
	&&& A1872 & 2009 Apr 30 & 4951.78337513 & 30 & 15930 \\ 
	&&& A1873 & 2009 May 01 & 4952.60611643 & 30 & 31260 \\ 
	&&& A2084 & 2010 Feb 16 & 5243.86334035 & 30 & 14460 \\
	&&& A2087 & 2010 Feb 17 & 5244.81192931 & 15 & 18645 \\ 
	&&& A2114 & 2010 Apr 10 & 5296.78896454 & 30 & 16770 \\
	&&& A2118 & 2010 Apr 12 & 5298.84209127 & 15 & 12255 \\ 
	&&& A2121 & 2010 Apr 16 & 5302.78442819 & 15 & 16545 \\ 
	&&& A2417 & 2011 May 28 & 5709.61961836 & 15 & 14595 \\ [3pt] 
 \object{SDSS J161531.72+454322.5} & $19.69\pm 0.02$ & 
 	$-0.42\pm 0.04$ & A1859 & 2009 Apr 21 & 4942.88180933 & 30 & 9030 \\ 
(SDSS J1615+4543)	 & & & A1862 & 2009 Apr 22 & 4943.89882089 & 30 & 7290 \\ [3pt] 
\objectname[SDSS J220029.09-074121.6]{SDSS J220029.09$-$074121.6} & $17.76\pm 0.02$ &
	$-0.46\pm 0.03$ & A1777 & 2008 Oct 26 & 4765.59790082 & 10 & 12000 \\ 
(SDSS J2200$-$0741) & & & A1991 & 2009 Oct 15 & 5119.57789282 & 10 & 11580 \\ 
	&&& A1996 & 2009 Oct 16 & 5120.56450366 & 10 & 12250 \\ 
	&&& A1999 & 2009 Oct 17 & 5121.57835663 & 10 & 11670 \\  
	&&& A2012 & 2009 Oct 22 & 5126.55784569 & 10 & 14460 \\ 
	&&& A2013 & 2009 Oct 23 & 5127.55688518 & 15 & 16980 \\ 
	&&& A2017 & 2009 Oct 24 & 5128.54691419 & 15 & 13620 \\ 
	&&& A2026 & 2009 Nov 11 & 5146.54809826 & 10 & 15510 \\ 
	&&& A2166 & 2010 Sep 09 & 5448.66327095 & 15 & 10140 \\ 
	&&& A2179 & 2010 Sep 13 & 5452.68328893 & 15 & 15555 \\ 
	&&& A2204 & 2010 Oct 09 & 5478.56723794 & 10 & 16900 \\ 
	&&& A2218 & 2010 Oct 12 & 5481.56595039 & 10 & 11510 \\ [3pt] 
 \objectname[SDSS J234843.31-094245.3]{SDSS J234843.31$-$094245.3} & $18.98\pm 0.03$ &
 	$-0.27\pm 0.04$ & A1782 & 2008 Oct 28 & 4767.64992309 & 30 & 16470 \\ 
(SDSS J2348$-$0942) & & & A1783 & 2008 Oct 29 & 4768.56183506 & 15 & 23475 \\ 
	&&& A1788 & 2008 Oct 30 & 4769.63453923 & 15 & 19485 \\ 
	&&& A1798 & 2008 Dec 25 & 4825.54142809 & 15 & 11340 \\ 
	&&& A1997 & 2009 Oct 16 & 5120.71354858 & 15 & 11535 \\ 
	&&& A2000 & 2009 Oct 17 & 5121.75860573 & 30 & 8400 \\ 
	&&& A2006 & 2009 Oct 19 & 5123.67006747 & 30 & 7290 \\ 
	&&& A2009 & 2009 Oct 20 & 5124.66921407 & 30 & 13590 \\ 
	&&& A2031 & 2009 Nov 13 & 5148.55165956 & 15 & 10125 \\ 
	&&& A2185 & 2010 Oct 02 & 5471.66983425 & 30 & 14460 \\ 
	&&& A2188 & 2010 Oct 03 & 5472.68240994 & 30 & 14430 \\ 
	&&& A2209 & 2010 Oct 10 & 5479.71038426 & 30 & 13560 \\ [3pt] 
\enddata
\tablenotetext{a}{Object names and photometry are from the SDSS DR7 White Dwarf Catalog\citep{2013ApJS..204....5K}; shortened names in parentheses are used in the remainder of this paper.}
\tablenotetext{b}{The start time is the mid-point of the first exposure.}
\end{deluxetable*}
\end{turnpage}
\clearpage
\global\pdfpageattr\expandafter{\the\pdfpageattr/Rotate 90}
\end{document}